\documentstyle[12pt]{article}
\def\s#1{{\bf#1}}
\def\tr{{\rm tr}}
\def\br{\hskip -0.18cm /}
\def\pr{{\it Phys. Rev.\ }}
\def\prl{{\it Phys. Rev. Lett.\ }}

\def\np{{\it Nucl. Phys.\ }}
\def\pl{{\it Phys. Lett.\ }}
\def\cmp{{\it Comm.Math. Phys.\ }}

\def\ap{{\it Ann. Phys.\ }}

\def\nc{{\it Nouvo Cim.\ }}
\def\aph{{\it Acta Phys. Hung.\ }}
\def\zf{{\it Z. Phys.\ }}
\newcommand{\mr}{\hline}
\newdimen\mathindent
\newcommand{\fl}{\hspace*{-\mathindent}}
\newcommand{\be}{\begin{equation}}
\newcommand{\ee}{\end{equation}}
\newcommand{\bea}{\begin{eqnarray}}
\newcommand{\eea}{\end{eqnarray}}

\begin{document}
\begin{center}
{\bf\LARGE Confinement and Renormalization
\footnote{Lecture delivered at the Enrico Fermi International School
of Physics on "Selected Topics in Non-Perturbative QCD",
June 1995, Varenna, Italy.}
}\\
\vspace*{2cm}
Janos Polonyi\\
\vspace*{1cm}
Laboratory of Theoretical Physics,\\
Louis Pasteur University, 3 rue de l'Universit\'e, 67084
Strasbourg, Cedex, France\\
\vspace*{.5cm}
Department of Atomic Physics, \\
Lorand E\"otv\"os University, 5-7 Puskin utca, Budapest 1088, Hungary\\
\vspace*{1cm}
{\LARGE Abstract}
\end{center}
The haaron gas description is reviewed for the QCD vacuum. The role of
non-renormalizable operators is emphasised in the mechanism which
generates the string tension. Additional examples are mentioned where certain
non-renormalizable operators of the bare lagrangian turn out to be
important at finite energy scale.
\vspace*{2cm}
\section{Introduction}
The confinement of quarks is a rather mysterious phenomenon of high energy
physics which challenges our understanding of Quantum Field Theory
and provides an ever reviving source of inspiration.
One may distinguish two different confinement mechanisms. The soft one,
which is responsible for the screening of the color charge of an isolated
quark by the creation of mesons \cite{gribov} in a manner reminiscent of the
sparkling of the supercritical vacuum of QED around a highly charged ion
\cite{gmr}. This latter occurs
when a single electron level dives into the Dirac see of negative energy
states. The electron of a virtual $e^+e^-$ pair fills up this hole and
the positron escapes to the infinity. The semiclassical condition for
this to happen is $Z\alpha_{QED}\approx1$. For hadrons $Z=2$ or 3 so one needs
a non-perturbative gluonic effect, the anti-screening, to raise the running
coupling constant to $\alpha_{QCD}\approx1$ at the confinement radius. This
confinement mechanism is called soft because it involves energies at the
range of the quark rest mass.

It is widely believed that the gluonic vacuum not only
amplifies the coupling constant at long distances according to the scenario
above but produces another phenomenon, the hard confinement mechanism.
This is the emergence of the linear string tension which leads to the
separation independent force between static quark charges. It
is a hard mechanism because the energy stored in the flux tube is
large for well separated charges.

The two mechanisms compete and it is very difficult
to disentangle them in the real world which contains the
virtual quark-anti quark vacuum polarizations. Only
the linear Regge trajectory provides circumstantial evidences for the
string picture. The more convincing proof of the hard confining mechanism
comes from lattice QCD.

It is clear that both mechanisms are needed to understand the problem
of quark confinement. But it is the anti-screening and the emergence of the
string tension in the gluonic vacuum which should be
clarified first beacause it is simpler. The soft confining mechanism can only
be discussed in the presence of such a non-perturbative medium. Our attention
will be limited on the simpler, hard mechanism in the rest of these lectures.

I think that the well defined and clean environment of
the numerical simulations of lattice QCD at finite temperature
where the deconfinement phase transition can be analyzed
is the ideal testing ground
for our ideas about confinement. It is mainly due to
the result of such numerical studies that the appropriate
kinematical framework and the dynamical content of the
phase transition can be identified. It was found that the
center symmetry which expresses the invariance of the gluonic
system under the fundamental group transformations is responsible
for the string tension \cite{mech}. The dynamical breakdown of this
symmetry by the kinetic energy at short time processes links
the non-perturbative vacuum with the asymptotically
free short distance phenomena and explains some rather unusual
features of the high temperature phase \cite{actphh}.

The invariance under the center or the fundamental group transformations might
be called "top-secret" symmetry after Coleman's beautiful Erice lecture
\cite{coleman} because it refers not only to the unobservable
gauge transformations but among them
to those which are represented trivially even on the gauge field.
Only the quark field transforms non-trivially under these transformations.
These transformations are mysterious because it is difficult to
isolate them in the continuum, can be broken dynamically without
loosing the global gauge invariance and are not protected by Ward-identities.
Another characteristic feature of this symmetry is that it requires nontrivial
measure in the functional integral which is invisible in dimensional
regularization. Furthermore the typical configurations in the symmetrical
realization of the path integral contain Dirac delta-type singularities.

I attempt to reconcile these unusual, sometime confusing features
of the hard confinement mechanism and perturbative QCD in these lectures.
We are far from the complete
quantitative solution and its outline, the haaron-gas vacuum \cite{laurent},
can be given only. The starting point is the
gauge invariance, the claim that the hard confinement mechanism
can be seen only when the gauge invariance is guaranteed exactly.
In fact, a non-controlled gauge dependent component of the vacuum state
represents a background charge which can screen the well separated
test quarks as the virtual quark-anti quark pair polarizations do it in
the complete vacuum. The ordinary gauge invariance is easier to guarantee
by satisfying the Ward-identities. The question of the center symmetry
is more subtle because it is not protected by simple identities
and can completely be missed in the usual dimensional regularization.
Thus we shall make sure that the center symmetry is present
not only formally but dynamically in the low energy effective theory
of the vacuum where the confining forces can be identified.
An interesting question we find in developing this effective theory,
namely the role of non-renormalizable operators will be discussed
in some details in the second part of the lectures.

\section{Effective theory of confinement}
This Section is devoted to the isolation and the characterization
of the mechanism which is responsible for the linear string
tension and the chiral symmetry breaking in the gluonic
vacuum. The center symmetry of the vacuum and its consequence is
discussed in the first part. A global symmetry does not give
a detailed enough picture of the dynamics so we need a local
effective theory for the order parameter. This is introduced
in the second part.

\subsection{Center symmetry}
\subsubsection{Functional Schr\"odinger representation}
The center symmetry is simplest to understand in the functional
Schr\"odinger representation. The canonical coordinate is the
gluon field, $\s A(\s x)=g\s A^a(\s x){\lambda^a\over2i}$, and the
corresponding momentum is the electric field,
$\s E(\s x)=\s E^a(\s x){\lambda^a\over2ig}$, where
$\s E^a(\s x)={1\over i}{\delta\over\delta\s A^a(\s x)}$.
The time component of the gluon field is eliminated by the choice of the
temporal gauge, $A_0(\s x)=0$ and the hamiltonian is of the form
\be
H=-2\tr\int d^3x\biggl[{g^2\over2}\s E^2(\s x)+{1\over2g^2}\s B^2(\s x)
\biggr],
\ee
where
\be
B_i\epsilon_{ijk}=\partial_jA_k-\partial_kA_j+[A_j,A_k].
\ee

The gauge fixing and the dynamics are invariant under
static gauge transformations,
\be
A_\mu(x)\to A_\mu^\omega(x)=\omega(x)(\partial_\mu+A_\mu)\omega^\dagger(x),
\ee
where $\omega(x)=\omega(\s x)\in SU(3)$. We are interested in the
gauge invariant vacuum sector where the propagator satisfies
\be
<\s A_f|e^{-itH}|\s A_i>_0=<\s A^\omega_f|e^{-itH}|\s A_i>_0. \label{ginv}
\ee
The usual way to select the gauge invariant contributions from
the general propagator $<\s A_f|e^{-itH}|\s A_i>$ is to insert a
projection operator into the vacuum sector,
\be
\fl<\s A_f|e^{-itH}|\s A_i>_0
=<\s A_f|{\cal P}_0e^{-itH}|\s A_i>
=\int D[\omega(\s x)]<\s A^\omega_f|e^{-itH}|\s A_i>,\label{vacpr}
\ee
where $\omega=e^{i\alpha^a\lambda^a}$ and
\be
{\cal P}_0=\int D_H[\alpha(\s x)]e^{i\int d^3x\alpha^a(\s x)\s D\s E^a(\s x)}.
\ee
The exponent contains the generator of the gauge transformations,
\be
\s D\s E=\partial\s E+[\s A,\s E]
\ee
and the integral variable appears as a static temporal component of the
gauge field when the path integral representation is worked out for
(\ref{vacpr}), $\alpha^a(\s x)=tgA_0^a(\s x)$,
\be
<\s A_f|e^{-itH}|\s A_i>_0=\int D_H[tgA_0(\s x)]\int D[\s A(\s
x,t)]e^{-S_{YM}},
\label{statpi}
\ee
with
\be
S_{YM}={1\over2g^2}\tr\int dx(\partial_\mu A_\nu-\partial_\nu A_\mu
+[A_\mu,A_\nu])^2.
\ee
When the projection operator ${\cal P}_0$ is inserted at each time slice,
$t_n=na$, then
the corresponding integral variable becomes the time dependent temporal
component of the gauge field, $\alpha^b(\s x)=agA_0^b(\s x)$,
\be
<\s A|e^{-itH}|\s A'>_0=\int D_H[agA_0(\s x)]\int D[\s A(\s x,t)]e^{-S_L}.
\label{pi}
\ee
We shall see below that the action is given in lattice regularization
is this case.

Global gauge transformations act as the basis transformations on the
gauge field,
\be
\s A(\s x)\to\omega\s A(\s x)\omega^\dagger.\label{center}
\ee
The center of the gauge group consists of those elements which commute with the
whole group, it is $Z_N=\{e^{i{2\pi\over3}\ell},\ell=1,\cdots,N\}$ for $SU(N)$.
Since the center element commute with the generators as well they leave the
gauge field invariant in (\ref{center}). In SU(2) gauge theory where the
center is $Z_2=\{1,-1\}$ and (\ref{center}) expresses the well known fact that
rotations by $2\pi$ leave the vectors invariant. The relation between the
center and the fundamental group can be seen by noting that the space of
global gauge transformations, $SU(N)/Z_N$, is
$N$-fold connected. The center symmetry is the invariance of the
propagator $<\s A_f|e^{-itH}|\s A_i>_0$
under global center transformations, (\ref{ginv}), with
$\omega(\s x)=e^{i{2\pi\over N}}$.

It is easy to construct an order parameter for the center symmetry.
The starting point is to note that the gauge invariance of
the path integral (\ref{pi}) is violated at the initial and the
final time slices. In fact, the boundary conditions
\bea
\s A(\s x,0)&=\s A_i(\s x)\nonumber\\
\s A(\s x,t)&=\s A_f(\s x)\label{bct}
\eea
do not allow to perform gauge transformations on the time boundaries.
The remaining symmetries at the initial and the final
time slice are the periodic gauge transformations with period length $t$
in time since the hamiltonian is gauge invariant, $[H,\s D\s E(\s x)]=0$.
Thus we have no Ward-identities for non-periodic gauge transformations.

The restriction of the space of gauge transformations to the periodic
ones increases the family of gauge independent variables \cite{tall}.
The observables generated in this manner correspond to the Polyakov line,
\be
\Omega(\s x)=Pe^{\int_0^tdt'A_0(\s x,t')},\label{poly}
\ee
which is the path ordered exponential along the stright line connecting
identical three-space points of the initial and the final time slices.
Its eigenvalues are invariant under the gauge transformations which are allowed
by the boundary conditions, (\ref{bct}). The gauge invariant eigenvalues,
$\lambda_a$, may serve as an order parameter because they transform
multiplicatively under center transformations,
\be
\lambda_a\to e^{i{2\pi\over N}\ell}\lambda_a.
\ee
Since the eigenvalues are not distinguishable our order parameter
will be their sum,
\be
\tr\Omega(\s x)\to e^{i{2\pi\over N}\ell}\tr\Omega(\s x).
\ee
It detects the center transformations which are not displayed by the
initial and final gluon fields, (\ref{bct}).

The insertion of the projection operator ${\cal P}_0$ into
the amplitude (\ref{vacpr}) makes the distribution of the order parameter
formally center symmetrical. But the dynamical preservation of the center
symmetry becomes energetically less favorable for short time processes.
This can be seeby the inspection of the effective action for
the Polyakov line \cite{effordp}, \cite{kornel},
\be
S_{eff}[\Omega]=-\ln<\s A^\Omega_f|e^{-itH}|\s A_i>.
\ee
It is formally center symmetrical but the potential barrier between
the center symmetrical minima diverges as $t\to0$ due to the large
kinetic energy needed for the finite global rotation of the gauge field
in short time \cite{bodrum}.

\subsubsection{Dimensional v.s. lattice regularization}
The Haar measure is defined by its invariance under group multiplication,
\be
\int d_H\omega f(\omega)=\int d_H\omega f(\omega'\omega).
\ee
This property is required in proving the gauge invariance of (\ref{vacpr}),
\bea
<\s A^{\omega'}|e^{-itH}|\s A'>_0
&=\int D_H[\omega]<\s A^{\omega'\omega}|e^{-itH}|\s A'>\nonumber\\
&=\int D_H[\omega]<\s A^\omega|e^{-itH}|\s A'>\nonumber\\
&=<\s A|e^{-itH}|\s A'>_0.
\eea

The invariant measure in the path integral (\ref{statpi})
can be taken into account perturbatively. For this end we write the
gauge transformation as $\omega=vhv^\dagger$ where $h$ is diagonal,
$h_{jk}=\delta_{jk}e^{iu_j}$. The Haar measure reads as
\be
d_H\omega=dvd^N\rho\sum_{n=\infty}^\infty\delta(\sum_ju_j-2\pi n)
\prod_{j<k}\sin^2{u_j-u_k\over2}.
\ee
For the sake of simplicity
we shall restrict ourselves to $SU(2)$ gauge theory.
Then $\omega=e^{u\hat n^j\sigma^j/2i}$ with $\hat n^2=1$ and
\be
d_H\omega=d\hat nd\rho\sin^2{u\over2}
=d^3u{1\over u^2}\sin^2{u\over2},
\ee
where $d\hat n$ is the uniform integration over $S_2$.
The Haar measure is a rotational invariant deformation of the
flat integration measure, $d^3u$.

We now return to the path integral where the projection operator is
inserted at each time slice and the argument of the Haar measure
is $agA^b_0(x)$,
\be
D_H[agA_0^b]=D[\hat n]D[u]e^{{1\over a^4}\int d^4x\ln\sin^2au(x)/2}.
\label{mvert}
\ee
Another form is where the Cartesian coordinate system is kept,
\bea
D_H[agA_0^b]&=D[A_0^b]
e^{{1\over a^4}\int d^4x\ln{1\over a^2u^2}\sin^2au(x)/2}\nonumber\\
&=D[A_0^b]e^{\int d^4x(-{u^2(x)\over12a^2}+{u^4(x)\over192}+O(a^2))},
\label{pmvert}
\eea
with $u^2=a^2g^2A_0^bA_0^b$ is better suited for perturbation
expansion. The UV divergent factor, $1\over a^4$, is needed
in order to make the exponent dimensionless. The nontrivial
integral measure can be taken into account by an additional potential
for the fields in local regularization schemes where the
elementary volume corresponding to a gauge degree of freedom can be identified.

The vertices introduced by the integral measure pose a new problem.
In our effort to implement the center symmetry we introduced a term
$O(u^2)$ which breaks the gauge invariance of the theory. The remedy
for this problem actually comes from another $O(u^2)$ piece in the action.
As was mentioned above the measure term can be treated consistently in
lattice regularization only. But the gauge invariance requires the use of the
link variables $U_\mu(x)=e^{aA_\mu(x)}$ instead of the gauge field,
$A_\mu(x)$. The expansion of the
plaquette in $A_\mu(x)$ gives rise another quadratic piece which cancels
the measure contributions and restores the gauge invariance at one-loop
level.

Similar cancellations between the lattice vertices continue to occur
at higher order. A formal proof of renormalizability by induction in
the order of the loop expansion can be given
by the help of the Ward-identities \cite{ymren}.
There are two kinds of genuine lattice vertices which have no analogy
in dimensionally regulated perturbative QCD. One, like the measure term,
is proportional to a negative power of the lattice spacing and UV
divergent. The other type which is suppressed by a positive power of the
lattice spacing. Neither of these vertices is problematical as far as the
overall
degree of divergence is concerned. In fact, simple power counting shows
that there are no new types of overall divergences because the dimension
of these new vertices is supplied by the cut-off itself. To show this
consider an observable computed in perturbation expansion where we have
one coupling constant only for simplicity,
\be
O=\sum_{n=0}^\infty g^nI_n,
\ee
where $I_n$ stands for a loop integral. Comparing the mass dimensions
of both sides we get $[O]=n[g]+[I_n]$. Since the overall degree of
divergence of the integral is $\omega(I)=[I]$ we have
\be
\omega(I_n)=[O]-n[g].
\ee
This relation shows that the renormalizable coupling constants must have
nonnegative dimension except when their negative dimension is provided
by the cut-off itself. The difficult
part of the proof is to show that the overlapping divergences are
removed as well and the counterterms can be chosen to be gauge invariant.
The final result is that the contributions of the diverging lattice vertices
cancel in each finite order of the loop expansion at high energy.
Furthermore, the order of the loop integration and the removal of
the cut-off, $a\to0$, can be exchanged since the properly regulated
theory contains finite, uniformly convergent loop integrals. It is
worthwhile noting that this holds only for theories without anomalies
\cite{rieszf}, \cite{rgqm}. Thus the contributions of the lattice vertices
cancel and the asymptotically free perturbation expansion based on the
usual three and four gluon vertices is recovered.

Dimensional regularization seems to be superior to the lattice regularization
because it skips from the very beginning those vertices whose contributions
are ultimately canceled. In fact, the suppressed lattice vertices are
dropped because only $1\over\epsilon$ or $\epsilon$-independent
pieces show up in the analytical regularization. The diverging
vertices are absent as well because
\be
{1\over a^4}=\int{d^4p\over(2\pi)^2}={\Omega_4\over(2\pi)^4}\int dpp^3
\ee
and the right hand side is set to zero. Are the complications of the lattice
regularization unnecessary ? They are certainly unimportant at high
energy and in any finite order of the loop expansion. But the situation changes
when resummation or non-perturbative approximation is sought. The periodic
Haar measure is essential to achieve the vanishing of the Polyakov
line and confinement at low temperature \cite{kornel}. For any
truncation of the Taylor expansion for the periodic potential in (\ref{mvert})
misses the periodicity and with it the confining forces. The perturbative
cancellation of the genuine lattice vertices is not enough to expel
them from the non-perturbative solution. But even if we had a
non-perturbative argument for the possibility of renormalizing the theory
in a gauge invariant manner it would not be enough to justify the neglecting of
the
genuine lattice vertices at finite energy where the asymptotic scaling
laws do not apply.

\subsubsection{Singular configurations}
Center symmetry implies that the eigenvalues of the Polyakov line,
(\ref{poly}), are distributed equally around the $N$-th roots of 1.
The local gauge invariance
of (\ref{pi}) allows us to set the integral variables $e^{agA_0(\s x,t)}$
to the unit matrix everywhere in the space-time except an arbitrarily
chosen equal time hypersurface where obviously we have
\be
e^{agA_0(\s x,t_0)}=\omega(\s x)\Omega(\s x)\omega^\dagger(\s x).
\ee
Whenever an eigenvalue of the Polyakov line is close to a nontrivial
$N$-th root of 1 we must have $agA_0=O(a^0)$. Thus the fraction ${N-1\over N}$
of the field configurations contain a Dirac delta type singularity,
$A_0=O(1/ag)$ in the continuum limit. In this manner
the apparently harmless observation
that the three vectors remains invariant under rotation by $2\pi$ brings badly
singular configurations in the renormalized path integral.
Since the projection operator ${\cal P}_0$ is inserted at each time slice
in (\ref{pi}) the configurations display the singularities in the fully gauge
invariant path integral, too. The naive continuum
expression for kinetic energy in the Yang-Mills lagrangian,
$-\tr(\partial_0\s A+\s DA_0)^2$ receives the term $\s DA_0$ from the
infinitesimal gauge transformations performed between the consecutive
time slices, $\delta\s A=a\s DA_0$. This expression is not invariant any more
when the gauge transformations are in a finite, cut-off independent distance
from the identity. The more careful derivation of the path integral expression
which takes such a singular configurations into account yields the usual
lattice regulated kinetic energy in terms of the link variable
$e^{agA_0(\s x,t)}$. The Lorentz invariant extension of the lagrangian
and the path integral leads unambiguously to Wilson's lattice gauge theory.

The singular structure of the configurations in the path integral is
the rule rather than an exception. In fact, consider the path integral
for a free massless particle in D dimension,
\be
\prod_x\int d\phi(x)e^{-{1\over2}a^{D-2}\sum(\phi(x+\mu)-\phi(x))^2}.
\ee
The typical configuration is where each contribution to the kinetic
energy is $O(a^0)$,
\be
\phi(x+\mu)-\phi(x)=O(a^{1-D/2}).
\ee
The trajectories are non-differentiable for $D=1$, in Quantum Mechanics,
\cite{schulm}, \cite{rgqm}, have finite discontinuities in two dimensions,
\cite{kerson}, and develop Dirac delta type power singularities for $D>2$.
The Fourier transform of the one-loop momentum space UV divergences yields
the same result. Due to this singular structure the topological concepts
introduced on the tree level may not survive the renormalization
procedure \cite{topren}.

\subsection{Effective theory}
\subsubsection{Sine-Gordon model}
Our goal is to construct a local effective theory which comprises the low
energy
effects of the center symmetrical fluctuations of the vacuum \cite{laurent}.
In the lattice regulated theory we find the usual three and four gluon
vertices, infinitely many genuine lattice vertices and the lattice propagator.
The lattice vertices cancel against each other perturbatively in the UV region.
We suspect that the periodicity of the integral measure is crucial at low
energies where the cancellation does not hold any more. Thus we shall ignore
all but the integral measure vertices for $A_0$.

The construction of the effective theory for SU(2) gauge theory
is as follows: We start in the gauge where $A_0$ is diagonal,
$A^a_0(x)=\delta^{a3}u(x)$ and a U(1) local gauge symmetry is left only.
Then we eliminate the off
diagonal gauge field components which are charged with respect to the
diagonal U(1) gauge group. The resulting effective theory is a nonlinear
U(1) gauge theory. So far this is the usual procedure for setting up
the Abelian confinement scenario, \cite{abelconf}, except that the Polyakov
line is used to single out the Abelian subgroup. Then we choose Feynman gauge
and eliminate the space-like component of the gauge field. We arrive at an
effective theory for $u(x)$ where the interactions come from
the effective vertices generated by the elimination procedure and
the measure term.

Is the measure term is the same in the effective
theory as in the original bare lagrangian ?
Recall that the boundary condition in time for the
trajectories in the path integral (\ref{pi}) restricts the gauge invariance
for periodic gauge transformations. Thus the integral measure for
the eigenvalues of the Polyakov line is not protected by gauge symmetry
and gets renormalized. But in the same time the center symmetry remains
present and requires that the integral measure be periodic. So the
integral measure for the diagonal component of
the local $A_0(\s x,t)$ must get renormalized as well. In fact, the eigenvalues
of the Polyakov line are given by the field $u(x)$,
\be
\lambda_a(\s x)=e^{\pm{ig\over2}\int dtu(\s x,t)},
\ee
and its integral measure would not be renormalized unless the
integral measure for $u(x)$ is renormalized.

In the spirit of the gradient expansion we ignore all derivative
coupling and truncate the effective theory onto the lagrangian
\be
L={1\over2}(\partial_\mu u)^2-V(u)\label{effl}
\ee
where the local potential is periodic,
\be
V(u+2\pi/\kappa)=V(u),
\ee
\be
V(u)=2\sum_{m>0}v_m\cos m\kappa u.
\ee
The contribution $m=0$ is eliminated by the condition $\int du V(u)=0$.
This four dimensional sine-Gordon type model is our effective theory.
It will be shown that despite the absence of the other interaction
it reproduces the salient non-perturbative features of the
vacuum.

\subsubsection{Haaron gas}
Consider first a simplified version of the effective theory with
one Fourier mode only,
\be
L={1\over2}(\partial_\mu u)^2-2\lambda\cos\kappa u.
\ee
The generator functional defined as
\be
Z[J]=\int D[u]e^{-\int d^4x[{1\over2}(\partial_\mu u(x))^2-\lambda(
e^{i\kappa u(x)}+e^{-i\kappa u(x)})+igJ(x)u(x)]},
\ee
what is expanded in $\lambda$,
\bea
&Z[J]&=\sum_{n=0}^\infty{\lambda^n\over n!}\prod_{j=1}^n
\bigl(\int d^4x_j\sum_{m(j)=\pm1}\bigr)\nonumber\\
&&\int D[u]e^{-\int d^4x[{1\over2}(\partial_\mu u(x))^2+igJ(x)u(x)]
+i\kappa\sum_{j=1}^nm(j)u(x_j)}\nonumber\\
&&=\sum_{n=0}^\infty{\lambda^n\over n!}\prod_{j=1}^n
\bigl(\int d^4x_j\sum_{m(j)=\pm1}\bigr)I_{mm}I_{JJ}I_{mJ},
\eea
where
\bea
&I_{mm}&=e^{-{\kappa^2\over2}\sum_{k,j=1}^nm(j)G(x_j-x_k)m(k)},\nonumber\\
&I_{JJ}&=e^{-{g^2\over2}\int d^4xd^4yJ(x)G(x-y)J(y)},\nonumber\\
&I_{mJ}&=e^{-g\kappa\sum_{j=1}^nm(j)\int d^4yG(x_j-y)J(y)},
\eea
and
\be
G(x)=\int{d^4p\over(2\pi)^4}{e^{ipx}\over p^2}={1\over4\pi^2x^2}.\label{coul}
\ee

The interpretation of these expressions is quite straightforward:
The sine-Gordon model is equivalent with the grand canonical ensemble
of point particles with charge $\pm\kappa$ and fugacity $\lambda$
interacting via the Coulomb potential (\ref{coul}). The logarithm
of the factors $I_{mm}$,
$I_{JJ}$ and $I_{mJ}$ contains the energies due to the self interaction of
the Coulomb particles, the source and the energy of the source in the
Coulomb fields of the particles, respectively.

We now return to the effective theory (\ref{effl}) where the repetition
the steps shown above gives
\bea
&Z[J]&=\int D[u]e^{-\int d^4x[{1\over2}(\partial_\mu u(x))^2+igJ(x)u(x)]
+2\sum_{m>0}v_m\cos m\kappa u}\nonumber\\
&&=\sum_{n=0}^\infty{1\over n!}\prod_{j=1}^n
\bigl(\int d^4x_j\sum_{m(j)=-\infty}^\infty v_{m(j)}\bigr)
I_{mm}I_{JJ}I_{mJ}.\label{haarp}
\eea
The charge is quantized in the units of $\kappa$ and the
Fourier coefficients turn into the fugacities. The Coulomb
particles are called haarons because they represent the effects
of the Haar measure.

The haaron-haaron interactions polarize the gas and introduce
the Thomas-Fermi screening. The screening mass square is the curvature
of the effective potential of the field $u(x)$ at the vacuum expectation
value. The manifest center symmetry, the invariance of the theory
with respect to the shift $u(x)\to u(x)+2\pi/\kappa$ allows the constant
effective potential only. This is in agreement with earlier numerical
finding, namely that the eigenvalues of the Polyakov line decouple from the
dynamics in the low temperature phase of QCD and its distribution is given
solely by the integration measure without the influence of the action
\cite{bill}. The absence of the screening in a Coulomb gas is attributed
to the negative fugacities. Some of the Fourier coefficients, $v_m$,
can be negative indicating negative probability in the classical Coulomb gas
picture of the sine-Gordon model. This makes the cancellation between the
screening effects of different charges possible. In other words,
the self energy computed in the framework of the perturbation expansion
may be vanishing for appropriately chosen coupling constants, $v_m$.
The dynamical issue of confinement is the assumption of the center symmetry,
the absence of screening in the effective theory. By the help of the
massless propagator, (\ref{coul}), the confining forces and the
chiral symmetry breaking can be obtained in a partial resummation of the
perturbation expansion.

After having excluded the generation of the screening mass the polarization
of the Coulomb gas becomes less important and the approximation $I_{mm}=1$,
the neglect of the Coulomb interactions between the particles is more
reasonable. The remaining contributions can be resummed in (\ref{haarp})
yielding
\bea
&Z[J]&=e^{-{g^2\over2}\int d^4xd^4yJ(x)G(x-y)J(y)
-\int d^4xV(ig\int d^4yG(x-y)J(y))}\nonumber\\
&&=e^{-\int d^4x({g\over2}J(x)U(x)+V(iU(x))}.\label{resum}
\eea
The exponent of the second equation
gives the energy of the source system
as the sum of the perturbative current-current interaction energy
and the potential energy in terms of the Coulomb potential created by
the source,
\be
U(x)=g\int d^4rG(x-y)J(y).\label{source}
\ee
The only unusual detail is the factor $i$ in the argument of the
potential $V$ whose role is to remove the periodicity after the
elimination of the quantum fluctuations. It is worthwhile noting that the
Wick rotation into real time is nontrivial and the factor $i$ remains
present.

\subsubsection{Static charges}
The energy $E$ of a test charge can be read off in the long time limit as
$tE=-\ln Z$ by choosing
\be
J(x)=\int ds{dy^0\over ds}\delta^{(4)}(x-y(s)),
\ee
where $y^\mu(s)$ is the world line of the charge. We take a static charge,
$y^\mu(s)=(s,\s x_0)$, and use (\ref{resum}),
\be
E_{np}=\int d^3xV\biggl({ig\kappa\over4\pi|\s x-\s x_0|}\biggr)
\ee
for the non-perturbative second term in the exponent of (\ref{resum}).

This result is typical inasmuch as the argument of the potential $V(u)$
is small in the IR since the Coulomb
field created by the source approaches zero at the infinity for localized
source. Thus the infrared contribution to the energy is controlled by the
behavior of the potential around zero, $V(u)=V(0)+{u^2\over2}V''(0)+O(u^4).$
Hence we have
\be
E_{np}=-V''(0)\biggl({g\kappa\over4\pi}\biggr)^2\int d^3x{1\over x^2}
+{\rm const.}\label{staten}
\ee
The curvature of the bare potential should be negative, $V''(0)<0$,
in order to arrive at flat effective potential in the IR.
The linear infrared divergence of the integral indicates the absence
of localized charged among the asymptotic states of the effective theory.
Note that the Thomas-Fermi screening would regulate this integral and
localized charges would be allowed. Thus one expects the effective theory
to undergo a phase transition and develop screening at the deconfining
phase transition.

Similar computation allows us to derive the static potential between
test charges. To this end we take
\be
J(x)=\int ds[\delta^{(4)}(x-y(s))-\delta^{(4)}(x-y(s)-L)],
\ee
with $L=(0,\s L)$. The non-perturbative part of (\ref{resum}) gives
\bea
&E_{np}&=-\int d^3xV\biggl({ig\kappa\over4\pi|\s x|}
-{ig\kappa\over4\pi|\s x-\s L|}\biggr)\nonumber\\
&&=-V''(0){g^2\over8\pi}|\s L|+\sum_{n=0}^\infty c_n|\s L|^{-n}.
\eea
The leading infrared part is a linearly rising potential with the string
tension
\be
\sigma=-V''(0){g^2\over8\pi},
\ee
and the coefficients in the sub-leading pieces, $c_n$, are ultraviolet
divergent.

\subsubsection{Nambu-Jona-Lasinio model}
It is straightforward to include quark fields into the effective theory.
The resulting lagrangian is
\be
L={1\over2}(\partial_\mu u)^2-V(u)
-\bar\psi[i\partial\br-g\gamma_0\sigma^3u]\psi.
\ee
It is not Lorentz or gauge invariant since the diagonal component
of $A_0$ are kept only after making approximations. The elimination of the
field $u(x)$ in the free haaron gas approximation gives (\ref{resum}) except
the external source is replaced by $J(x)=\bar\psi\gamma_0\sigma^3\psi$,
\bea
&S&=\int d^4x\biggl\{-\bar\psi\partial\br\psi+
V(ig\int d^4yG(x-y)J(y))\biggr\}\nonumber\\
&&-{g^2\over2}\int d^4xd^4yJ(x)G(x-y)J(y)\nonumber\\
&&=\int d^4x\biggl\{-\bar\psi\partial\br\psi+
V(iU(x))-{g\over2}\int d^4xJ(x)U(x)\biggr\},
\eea
where (\ref{source}) is used to express the Coulomb field of the
dynamical source $J$. The second term in the first equation
gives an infinite series of non-local vertices. The $O(J^2)$ action is
\bea
&S&=-\int d^4x\bar\psi\partial\br\psi\nonumber\\
&&+\int d^4xd^4yJ(x)\biggl\{-{g^2\over2}G(x-y)
+{4\pi\sigma\over g^2}G_2(x-y)\biggr\}J(y),\label{njl}
\eea
where
\be
G_2(x)=\int{d^4p\over(2\pi)^4}{e^{ipx}\over p^4}.
\ee
Notice that the static potential generated by $G_2(x)$ is linear
with string tension $\sigma$. There have been several studies of the
NJL model with such kind of interaction and the self consistent gap equation
approximation shows the dynamical breakdown of the chiral symmetry induced
by the strong repulsion \cite{enjl}.

It is illuminating to eliminate the quark field degrees of freedom
in favor of a composite meson field. By the help of the identity
\be
\fl\int D[\Phi]e^{-{1\over4}\int dxdy\Phi(x)K^{-1}(x,y)\Phi(y)
+\int dxA(x)\Phi(x)}=e^{\int dxdyA(x)K(x,y)A(y)}\label{hubb}
\ee
we find
\be
S=-\tr\ln[i\partial\br+\gamma_0\sigma^3\Phi]
+{1\over2g^2}\int dx\Phi{\partial^4\over{4\pi\sigma\over g^2}+\partial^2}\Phi.
\label{enjle}
\ee
Notice the strong IR dependence of the wavefunction renormalization constant
\be
Z(p^2)={p^2\over{4\pi\sigma\over g^2}-p^2}.
\ee
Since the restoring force for the small fluctuations of the meson field is
weak in the IR there are strong interactions between the low energy mesons
and the quark-anti quark vacuum polarizations.

It is worthwhile noting that the Euclidean mesonic effective theory is
defined only below the Landau pole $p^2<{4\pi\over g^2}\sigma$ since beyond
this limit the Hubbard-Stratanovich transformation (\ref{hubb}) requires
imaginary meson field.

\subsubsection{Haaron gas and localization}
There is a simple physical picture behind the haaron gas description of the
QCD vacuum. Consider the quenched multi-quark Green function in the
free haaron gas approximation,
\bea
&<0|&T[\bar\psi(y_1)\cdots\bar\psi(y_n)\psi(z_1)\cdots\psi(z_n)]|0>\nonumber\\
&&=\sum_{n=0}^\infty{1\over n!}\prod_{j=1}^n
\bigl(\int d^4x_j\sum_{m(j)=-\infty}^\infty v_{m(j)}\bigr)\nonumber\\
&&<\bar\psi(y_1)\cdots\bar\psi(y_n)\psi(z_1)\cdots\psi(z_n)>_U,
\label{haarg}
\eea
where the factor in the last line is the free fermion Green function
on the imaginary background field
\be
U(x)={i\kappa\over4\pi^2}\sum_{j=1}^n{m(j)\over(x-x_j)^2},
\ee
e.g.
\be
<\bar\psi(y)\psi(z)>_U
=\biggl[{1\over i\partial\br-g\gamma_0\sigma^3U}\biggr](y,z).
\ee

The expression (\ref{haarg}) is reminiscent of the fermionic Green functions
in the dilute instanton gas approximation. But our result is not a
semiclassical contribution, the haaron gas may be dense and is not
an extremum of the Yang-Mills action. Furthermore there are no
zero modes and result is infrared finite. For the dilute haaron gas the fermion
propagator can be factorized and the zero mode dominance yields the
condensate
\be
<0|\bar\psi\psi|0>=-V(0)\tr\int d^4x<\bar\psi(x)\psi(x)>_U
\ee
where propagator in the right hand side is evaluated on a single haaron
background. Similarly to the case of the instantons there are localized
zero modes for the Dirac operator which make the integral finite and generate
the chiral symmetry breaking.

The hard confinement mechanism is similar to localization
observed in strongly correlated electron systems, \cite{anderson}, except that
it takes place in the space-time rather than the three-space. We have seen
that the IR divergence of the single quark energy makes the quark propagator
vanishing. It is easy to reproduce this result by the localization scenario.
In fact, consider an isolated quark which is sent through the vacuum. The
oscillation due to the phase shift which is generated by the long range
haaron potential cancels the propagator.

In order to understand the propagation of a quark-anti quark pair
we have to take into account the correlation between the haarons. The
coupling constants $v_m$ provide the scale for the finite haaron density
in the vacuum and the average haaron distance appears as a correlation
length of the quenched Coulomb potential generated by the haarons.
Let us follow a very simple way of keeping track of this correlation:
The haaron potential is considered completely correlated or uncorrelated
for separations which are less or more than the correlation length,
respectively. The propagator of a meson can be written as the sum
over the world lines of the quark-anti quark pairs.
The phase shifts of the quark-anti quark state cancel so long as the
separation of the quark-anti quark pair is smaller than the correlation length.
When the distance between the pair increases beyond the correlation length then
the statistically uncorrelated potential creates a non-vanishing
phase shift which in turn suppresses the contribution after the
haaron gas averaging. Thus the quark-anti quark pair tends to stick
together and the confinement radius is the correlation length of the
haaron potential. The similarity between this scenario and the realization
of confinement in the stochastic vacuum \cite{stoch} is remarkable.

\subsubsection{Double role of $V(u)$\label{double}}
The periodic potential $V(u)$ has two rather different role
in our approximate solution. On the one hand, it controls the large
amplitude fluctuations of the field $u(x)$ in the effective lagrangian,
(\ref{effl}). The order of magnitude of the center symmetrical fluctuations
is ${1\over\kappa}$ and the behavior of $V(u)$ in this range,
in particular the periodicity, is important. On the other hand,
the same potential appears after the partial resummation in the
free haaron gas. Its the typical argument is purely imaginary and small
in absolute magnitude as far as the IR physics is concerned because the
Coulomb potential approaches zero for large distances. The approximation
$V(u)\approx V(0)+V''(0)u^2/2$ around the origin is sufficient to express
the string tension and the chiral condensate. After the elimination
of the fluctuations the non-fluctuating argument of the potential is
small and the IR physics is governed by the behavior of the potential
around zero.

One finds that the periodicity of the potential
and the manifest center symmetry
has two consequences in the effective theory: In the
UV it keeps the fluctuations of the field $u(x)$ large and thereby
it creates the non-perturbative environment.
In the IR regime it protects against mass generation. This
latter appears as the discrete analog of the chiral symmetry.
What is furthermore interesting is that a discrete symmetry is responsible
for the massless behavior.

\subsubsection{Center symmetry and instantons}
The perturbative ordered vacuum is based on the configuration
$A_\mu=0$ and the small fluctuations around it. This gives
${1\over V}\int d^3x\tr\Omega(\s x)\approx2$ in $SU(2)$
theory which indicates that the transition
amplitudes are not center symmetrical. What modes are responsible for the
restoration of the center symmetrical vacuum,
${1\over V}\int d^3x\tr\Omega(\s x)=0$? It is reasonable to
expect that localized configurations which have large entropy
and interpolate between the center related minima of the
effective potential for the order parameter will
reach this goal for sufficiently large values of the time parameter $t$
of the matrix element of the time evolution operator (\ref{pi}).

The shape of such an
interpolating configuration can be found as the solution of the
Yang-Mills equations of motion. This turns out to be an instanton. The reason
is that the map of the three-space into the gauge group is $S_3\to S_3$ when
the space is spherically compactified. The Polyakov line configurations
are labeled by the Pontryagin index, the winding number. But this is just
the topological charge, \cite{tcltg}. Thus the Polyakov line of an instanton
winds around the $SU(2)$ group space as the space coordinates
moves around the whole three-space. The asymptotical value for large
coordinates
is $\Omega\approx1$ and in order to cover the whole group space it has  to
take the value $\Omega=-1$ somewhere. Due to the rotational symmetry
of the solution this happens at the center of the instanton. In this manner
$\tr\Omega(\s x)$ is a spherically symmetrical localized
function which interpolates between $\pm2$. The gas of such "domains" restores
the center symmetry.
Note that the size of the instantons which might play role in the center
symmetry restoration must be around the confinement radius.

\section{Universality and condensates}
The picture of the confining gluonic vacuum outlined above raises
more questions than answers. The justification of the emphasis put
on the center symmetry comes from the numerical experiences in lattice
gauge theory, namely from the observation that the fate of the center
symmetry is related to
the existence of the string tension. But the important IR effects
are produced by the Haar measure vertices of the path integral, by
the vertices which are labeled as non-renormalizable or irrelevant
according to the renormalization group. These vertices are
set to zero in dimensional regularization. Does that mean that
the usual class of renormalizable field theories is not sufficient
to parametrize all possible physics of a given set of particles and there might
be different QCDs? We are not in the position to answer affirmatively this
question which goes beyond the perturbation expansion. Instead, my goal will be
to indicate only a gap in the usual argument which might make possible to
describe the non-perturbative IR effects more systematically and in the
same time to find new continuum field theories.

There is a well known example where the usual power counting might miss
a relevant operator, the strong coupling QED \cite{miransky}. The point is
that the anomalous dimensions should be included into the usual power
counting argument in classifying the operators of the theory. The extrapolation
of the perturbative anomalous dimension of the electron field
gives dimension four for the four fermion operator when $e\approx1$.
Thus we have the possibility for a new relevant coupling constant
in strong coupling QED.

We shall give two other examples for the generation of new relevant
coupling constants. They might be more realistic because one of them
refers to asymptotically free theories and the other is the Higgs
sector of the Standard Model. In both cases the apparent violation of the
universality and the emergence of new parameters are related to
condensates.

\subsection{Localized saddle points}
Our example for non-perturbatively generated relevant coupling constants
in asymptotically free models is based on the higher order derivative terms
in the lagrangian. If localized saddle points, coherent states, appears
in the theory then the higher order derivatives may deform them substantially.
Depending on the sign of the derivatives the saddle points may shrink
to the cut-off size and saturate the path integral with cut-off effects
at each length scale.
What is interesting is that this may happen despite the weakness of the
coupling constant at the cut-off scale because the derivatives
of the saddle point are sufficiently large.

\subsubsection{Bare vs. renormalized expansion}
The starting point is the difference between the bare and the renormalized
saddle point expansion \cite{enzo}. The well defined path integral
is given only for the bare, regulated theory,
\bea
&\int D[A_\mu]e^{-S_B}&=\int D[A_\mu]e^{-S_0-S_i-S_{CT}}\nonumber\\
&&=\int D[A_\mu]e^{-S_0}\biggl\{1-S_i-S_{CT}+\cdots\biggr\},
\eea
where the small parameter of the perturbation expansion is $g_B$.
The bare action is the sum of the renormalized one, $S_R=S_0+S_i$,
and the counterterms, $S_{CT}$.
The new expansion parameter, $g_R$, is obtained in the renormalized
perturbation expansion after taking into account
the cancellations in the Taylor expansion of the second equation. But note
that this procedure is formal in the sense that there is no well defined
path integral with the small expansion parameter $g_R$. In such a manner
the reliability and the applicability of the perturbation expansion should
be investigated in the framework of the bare rather than the renormalized
series since the expansion is done before the cancellations take place.
In fact, the phase transitions provide evidences that the renormalizable
running coupling constants can not even characterize the theory in
a unique manner.

The difference between these two expansion schemes is more pronounced
for theories with dimensionless coupling constants only, such as the
$SU(2)$ Yang-Mills theory. Here the renormalized saddle point expansion
\cite{thoofti} is based on the strategy that the saddle points are selected
by $S_R$ and the counterterms are taken into account on the
higher loop level only. The tree level saddle points are the instantons
whose action is independent of the instanton size due to the scale invariance
of $S_R$. The higher loop contributions break this scale
invariance in a manner that large instantons are preferred and the
non-interacting instanton gas picture becomes inconsistent.
The bare saddle point expansion brakes the scale invariance on the tree
level already since a scale parameter, the cut-off, appears explicitly
in the regulated action. Such a tree level violation of the scale invariance
is governed by the dimensional parameters of the bare action. In order
to explore the possibilities of the breakdown of the scale invariance
we introduce additional dimensional coupling constants in the bare theory. The
coupling constants with positive mass dimension are superrenormalizable
and influence the distribution of the large, i.e. cut-off independent
saddle points. They are excluded in Yang-Mills theory by symmetry.
The coupling constants with negative mass dimension are non-renormalizable
and govern the distribution of the ultraviolet saddle points around the
scale of the cut-off.

The regulators are represented by irrelevant operators because they are
supposed to supress the fluctuaitons in the UV regime only. In this manner
different regulators give different distribution for the small instantons.
The instantons whose size parameter is in the vicinity of the cut-off and
whose dynamics is influenced by the regulators will be called mini-instantons.
The usual Wilson-type single plaquette action decreases
monotonically with the size of the instantons. The too low action of the
mini-instantons leads to divergent topological susceptibility and
nonunique topological charge in lattice regularization. The usual
strategy to cope with this problem is to construct topological charge
operator or improved action which cuts out or suppresses the mini-instantons,
alas topological defects \cite{topdef}. This is certainly justified
when we are to establish a lattice regulated Yang-Mills model
as close as possible to the continuum perturbation expansion.
But our goal is different in the search of the confinement mechanism
when the results of the dimensional regulated renormalized saddle
point expansion can not always be used. Instead, we take the bare
theory seriously, on a non-perturbative manner at $each$ length scale
and try to trace downs the modifications of the dynamics due to the
mini-instantons, if exist, even in the IR regime.

\subsubsection{Mini-instantons}
We start with the bare theory which involves some higher dimensional terms,
\be
L_B=-{1\over4g^2}F^a_{\mu\nu}\biggl(\delta^{ab}+{c_2\over\Lambda^2}D^{2ab}
+{c_4\over\Lambda^4}D^{4ab}\biggr)F^{a\mu\nu}.
\ee
The higher order pieces usually come from a heavy particle exchange
in the energy range $\Lambda$. The coupling constants $c_\alpha$ are
irrelevant according to the power counting.
The simplest way to see this is to consider an observable obtained in the
perturbation expansion,
\be
\fl<O>=\mu^{[O]}
O\bigl(g^2,c_2{\mu^2\over\Lambda^2},c_4{\mu^4\over\Lambda^4},{\mu\over\Lambda}
\bigr)=\mu^{[O]}\sum_{jk\ell}
g^j\biggl({c_2\mu^2\over\Lambda^2}\biggr)^k
\biggl({c_4\mu^4\over\Lambda^4}\biggr)^\ell I_{jk\ell}({\mu\over\Lambda}).
\label{pertir}
\ee
The characteristic scale, $\mu$, of the observable is used to give its
dimension and each insertion of a vertex with the coupling constant
${c_\alpha\over\Lambda^\alpha}$ brings the factor $\mu^\alpha$
by dimensional reasons. So long as the theory is infrared finite and the
limit ${\mu\over\Lambda}\to0$ is convergent the $c_\alpha$ dependence
drops in the renormalized observables.
The power counting gives another important result, which has already
been mentioned above in connection with the renormalizability of lattice
gauge theory. Namely, the coupling constants $c_\alpha$ do not
harm renormalizability, they are actually a variant of the
Pauli-Villars regulators.

The power counting argument is no longer valid if there are
localized saddle points in the theory. In that case the dependence
in the coupling constants is not necesseraly polynomial and the suppressing
factor $({\mu\over\Lambda})^\alpha$ may be missing. As an example take
$Z_1/Z_0$, the ratio of the partition function of the one and the zero
instanton sector. In order to find the saddle point
consider a one parameter family of the instanton configurations labeled
by a scale parameter $\rho$. The action of the instanton is
\be
S_B(\rho)=-{8\pi^2\over4g^2}\biggl(1-{\tilde c_2\over(\Lambda\rho)^2}
+{\tilde c_4\over(\Lambda\rho)^4}\biggr)
\ee
where $\tilde c_\alpha/c_\alpha,\ \alpha=2,4$ are positive, $\rho$ and $g$
independent constants. The stable mini-instanton size, $\tilde\rho$, is
obtained
by solving $(\tilde\rho\Lambda)^2=2\tilde c_4/\tilde c_2$,
\be
S_B(\tilde\rho)={8\pi^2\over g^2}\biggl(1-{\tilde c_2^2\over4\tilde
c_4}\biggr).
\label{minst}
\ee
The one loop approximation yields
\be
{Z_1\over Z_0}=Cg^pV\Lambda^4e^{-S_B(\tilde\rho)},\label{partrat}
\ee
where $V$ is the four volume \cite{enzoym}. Observe the absence of the
perturbative suppressing factor $1/(\Lambda\tilde\rho)^2$ in (\ref{minst})
and (\ref{partrat}). What is not suppressed here is a cut-off
contribution because $1/(\Lambda\tilde\rho)^2=O(\Lambda^0)$.

A more careful
analysis of the fluctuation determinant shows that the mini-instantons
dominate (\ref{partrat}) for $1-{\tilde c_2^2\over4\tilde c_4}<X<1$, where $X$
is a given positive constant. When this inequality is satisfied then one of the
following two possibilities is realized: (i) $c_\alpha$ are relevant or
(ii) $c_\alpha$ are irrelevant but the beta function for $g$ is non-universal.
In fact, let us assume that $c_\alpha$ are irrelevant. Then the cut-off
independence of (\ref{partrat}) gives the beta function
\be
\beta_g=\Lambda{dg\over d\Lambda}
=-{g^3\over4\pi^2(1-{\tilde c_2^2\over4\tilde c_4})}<-{g^3\over4\pi^2X}.
\ee
In either case the coupling constants $c_\alpha$ modify the physical
content of the theory at length scales which are independent of the cut-off.
In other words, the saturation of the path integral by the mini-instantons
which are close to the cut-off makes the cut-off effects present and changes
the dynamics at finite length scales.

\subsection{Multiple fixed points}
Another example for the unusual relevant parameters is when there are
several fixed points in a theory and
a given operator is relevant at one fixed point and irrelevant at
another one. The appearance of multiple fixed points is typical in Particle
Physics where different interactions are found to be dominant at
different energy scales.

\subsubsection{Theory of Everything}
Let us imagine the renormalized trajectory of the Theory of Everything (TOE).
This theory includes all physics, at any length scale, by definition.
In order to understand
its features it is useful to summarize the connection between the language
of the renormalization group when applied in Statistical Physics and in
Quantum Field Theory.

\begin{table}
\begin{tabular}{@{}*{2}{l}}
\mr
Statistical Physics&Quantum Field Theory\cr
\mr
U.V. fixed point&Renormalized theory\cr
Irrelevant coupling constant&Non-renormalizable coupling constant\cr
Relevant coupling constant&Renormalizable coupling constant\cr
Universality&Considering renormalizable theories only\cr
\mr
\end{tabular}
\end{table}

The UV fixed point is where the correlation length is infinite. It corresponds
to the infinite value of the cut-off, to the renormalized theories.
The applicability of the linearized version
of the blocking relations gives the scaling regimes. In these regions we
have a classification of the coupling constants. Those coupling
constants which decrease or increase when we move
towards the IR regime are called irrelevant or relevant, respectively.
The irrelevant coupling constants increase as
we move into the opposite, UV, direction. This is what we always do in the
renormalization of a Quantum Field Theory. Thus the irrelevant operators
prevent us from going "back" to the UV fixed point and from removing the
cut-off. Hence they are non-renormalizable.

Finally, universality states that the physics at the length scale
which corresponds to the IR end of the scaling regime is insensitive for the
choice of the irrelevant coupling constants. This translates into the
usual rule of model building in Particle Physics, the use of renormalizable
theories only. The non-renormalizable theories were first excluded
due to their uncontrollable UV behavior. By some tricks, such
as the use of the cut-off to suppress the non-renromalizable coupling
constants mentioned above may help to eliminate the divergences from the
theory but the predictive power and the simplicity are lost. A simpler
argument to ignore non-renormalizable coupling constants is offered by the
renormalization group: They become small anyhow at the scale of the
observations so we might as well set all of them zero from the very beginning,
at the cut-off.

We now return to the renormalized trajectory of the TOE.
At its asymptotically high energy scaling regime, in the vicinity of the
UV fixed point the trajectory is governed by the relevant coupling constants.
But it is the lesson of the Wilson-Kadanoff blocking that $all$ coupling
constants what might be generated later should be included into the
theory from the very beginning. Since the TOE describes the physics down
to the classical regime all effective
coupling constants must appear in the theory. In this manner
the coupling constant space includes GUT, Standard Model, Nuclear, Condensed
Matter, Solid State and Atomic Physics parameters, too. Say a QCD quark-gluon
vertex appears as a multi-particle composite vertex on the level of the TOE.
We must include all composite vertices in the action what might be needed
to characterize the physics at lower energies.

Suppose that we increase the energy of the measurement from few GeV.
The renormalized trajectory converges to the UV fixed point of QCD only
in the model computations. In the real world the electro-weak interactions
which are represented by the non-renromalizable current-current
vertices well below the Standard Model scale start to deflect the
trajectory form the QCD UV fixed point. As we further increase the
energy the local effective vertices explode at the threshold of the
weak vector bosons. From now on we see the scaling laws of the
Standard Model. The renormalized trajectory stays in the vicinity of the
UV fixed point of the Standard Model (the triviality problem of the
Higgs sector is now ignored for simplicity) so long as the effective vertices
generated by the exchanges of a superheavy vector boson are weak. By repeting
this argument at each intermediate fixed point we find that the
renormalized trajectory of the TOE visits
several fixed points, those of the GUT, Standard Model, QCD, QED,
Condensed Matter, Solid State and Atomic Physics among others, as we
move towards the IR. Finally there is an IR fixed point.
There are further fixed points in the coupling constant space which can
be approached by the renormalized trajectory when the environmental
parameters, such as the temperature or chemical potentials are properly
tuned.

Since the same operator algebra is classified
in the vicinity of each fixed point it may happen that a given operator
turns out to be relevant at some fixed point and irrelevant at others.
It is certainly right that the renormalized trajectory is influenced
only by the relevant operators of a fixed point in the scaling regime of
the fixed point in question. Thus we have islands of universality around each
fixed point. But a coupling constant which happens to be irrelevant
at one fixed point may turn out to important at another fixed point.
Thus it is a quite involved question
that what operators are relevant at a certain energy range because the
answer depends not only on the energy range considered but all scaling
regime between the energy range in question and the true UV fixed point.
We shall see below the possibility that other fixed points toward the IR
direction may influence the result, too.

\subsection{From the superconductor to the Standard Model}
To simplify the mixing of different fixed points
consider the strong and the electromagnetic interactions
for electrons, muons and nucleons at finite baryon density. This gives a
well defined UV scaling regime until the Landau pole of QED.
For the appropriate choice of the chemical potential we may have another
scaling regime at lower energy which is related to Condensed Matter or
Solid State Physics. The two scaling regimes define two classification schemes
for the operators. Let us call an operator relevant or irrelevant at a fixed
point if it is contained in the relevant scaling operator set of the fixed
point or not, respectively. Each operator belongs to one of the
classes (rel,rel), (rel,irr), (irr,rel) or (irr,irr).
Here the first and the second property refers to the behavior
of the operator in the UV or the IR scaling regime. The electron mass, $m_e$,
is of the type (rel,rel) since it is a renormalizable parameter of the
QED lagrangian and influences the phase transitions at lower energies.
On the contrary, the muon mass, $m_\mu$, is (rel,irr) because the effects of
the
muons are shielded by those of the electrons at energies below $m_\mu$.

The interesting class is the (irr,rel). A coupling constant of that
type drops as we lower the observational energy from the UV cut-off
and becomes undetectable. But it starts to grow as we arrive at the
IR scaling regime and may play important role there and at lower
energies. This is a "hidden parameter" in the sense that it must be
specified at the microscopical scales but its effects appear at much longer
length scales only. It represents an elementary interaction which is not
detectable on the level of the elementary constituents and becomes important
only when certain long range structure such as the solid state lattice is
formed.
An example of this type of operator is an effective four fermion
contact term which incorporates the weak attractive forces between the
electrons of the solid due to the phonon exchange. This non-renormalizable term
is responsible for the superconducting ground state whose influence
becomes dominant at very low energy. This four fermion contact term generated
by the phonons is the analogy of the contribution $O(J^2)$ in (\ref{njl}).

According to the strategy of the Wilson-Kadanoff blocking all operators
which are generated at lower energy scales should be included in the
hamiltonian from the very beginning. The appearance of non-renormalizable
terms should not be a serious problem except for the
UV fixed point of the TOE. In fact, since all lower energy
theory is effective only the scaling laws inferred in the vicinity of
a fixed point change as we go up in energy and the apparent explosion of the
non-renormalizable coupling constants gives rise to stability
in the inter-fixed point region. One is left with wild speculations only
concerning the TOE. A possible scenario to include
all coupling constant is to require the UV finiteness of this ultimate
theory.

The crucial question whose clarification requires the detailed
quantitative analysis is whether the initial value for the coupling constants
of the type (irr,rel) effect the infrared behavior of the theory. In other
words, whether the dynamical growth of these coupling constants as we
pass the crossover in lowering the cut-off between the two fixed points
is modified by the initial conditions taken at the UV side. In case
of QED this is the question whether the Fermi constant or other
effective vertex of similar structure generated at the level of the
Standard Model or beyond influences the supercurrent density
in solids.

We find an aspect of the condensate formation in the example above which is
different from those observed in regard to the mini-instantons.
The four fermion
interaction becomes important at energies which are well below the
characteristic mass scale, $m_e$. What kind of degrees of freedom
correspond to this new operator ? One part of the answer
is trivial. Namely, the degrees of freedom which are responsible
of this vertex belong to the coherent state of photons with nonzero momentum,
the solid state lattice. The excitations of this condensate, the phonons,
generate the effective vertex. The other part of the question, the degrees
of freedom influenced by this vertex is less obvious. They come from
a condensate again, the boose condensate of the Cooper pairs. Thus we have
two condensates below $m_e$ and their dynamics are related to a
non-renormalizable operator in the lagrangian. Other field theoretical
models show similar behavior \cite{af}.

Historically, the discovery of the superconducting state of matter was a
total surprise. The understanding of Quantum Field Theory
was not sufficient at that time to predict such phenomena by a theory
which was well tested in few particle process only. The renormalization
group provides us the language and the framework to study problems like
this and the systematic search of relevant but non-renormalizable
operators can reveal such surprises.

\subsubsection{Wegner-Haughton equation}
In order to separate the impact of different scaling regimes on the
renormalized trajectory one has to follow the mixing of each operator
which might prove to be relevant at some fixed point or important at
the crossowers. The differential form of the renormalization group equation,
\cite{weghau}, is more suited for this goal because it handles
the mixing of infinitely many coupling constants in a very economical manner.
We shall derive the equation for a scalar field theory by using sharp
momentum space cut-off.

Denote the bare action by $S_k[\phi]$ where $k$ is the UV cut-off. We shall
obtain $S_{k-\nabla k}[\phi]$ in the loop expansion,
\be
\fl e^{-S_{k-\nabla k}[\phi]}=\int D[\tilde\phi]e^{-S_k[\phi+\tilde\phi]}
=e^{-S_k[\phi+\tilde\phi_0]-{1\over2}\tr\ln
{\delta^2S[\phi+\tilde\phi_0]\over\delta\phi\delta\phi}+O(\nabla k/k)},
\label{wh}
\ee
where the Fourier amplitude of the fields $\phi(x)$ and $\tilde\phi(x)$
is non-vanishing for $p<k-\nabla k$ and $k-\nabla k<p<k$, respectively
and the saddle point is given by
${\delta S[\phi+\tilde\phi_0]\over\delta\phi}=0$. Since the n-loop
contributions include an n-fold integration over the functional space
$\tilde\phi$ they are proportional to $(\delta k/k)^n$ in the
absence of massless singularities in the given kinematical region.
Thus $\delta k/k$ appears as a new small parameter and the exact functional
differential equation obtained in the limit $\nabla k\to0$ includes the
one-loop contribution only.

In order to simplify (\ref{wh}) we use the gradient expansion,
\be
S[\phi]=\sum_{n=0}^\infty\int dxU_n(\phi(x),\partial^{2n}),\label{grex}
\ee
where $U_n$ is a homogeneous function of order $2n$ in the derivative.
This expansion will be truncated at $n=1$,
\be
S[\phi]=\int dx{Z(\phi)\over2}(\partial\phi)^2+U(\phi),
\ee
and furthermore the simplification $Z(\phi)=1$ will be used to derive
a simple differential equation for the potential $U$. In order to
pick up the local potential from the action we choose a homogeneous background
field $\phi(x)=\Phi$. The functional differential equation reduces to
\be
e^{-VU_{k-\nabla k}(\Phi)}=e^{-VU_k(\Phi)-{V\over2}\int{d^dp\over(2\pi)^d}
\ln(p^2+V_k''(\Phi))},
\ee
where the $V$ stands for the space-time volume and the
integration extends over the shell $k-\nabla k<p<k$. In the
limit $\nabla k\to0$ one easily finds
\be
k\partial_kU_k(\Phi)=-{\Omega_dk^d\over2(2\pi)^d}\ln(k^2+U_k''(\Phi))
\label{drge}
\ee
where $\Omega_d$ denotes the d-dimensional solid angle.
This equation represents the one-loop resummed mixing of the coupling
constants of the potential
$U_k(\Phi)=\sum_ng_n\Phi^n$. In fact, the expansion of the logarithm in
the second derivative of the potential gives
\be
k\partial_kU_k(\Phi)=-{\Omega_dk^d\over2(2\pi)^d}\sum_n
{1\over n}\biggl({-U_k''(\Phi)\over k^2+U_k''(\Phi)}\biggr)^n,
\ee
up to a field independent constant. This is the sum over
the Feynman graphs contributions which come from the
infinitesimal loop integration volume. The circumstance that the right
hand side includes the running potential $U_k(\Phi)$ rather than the
bare one, $U_\Lambda(\Phi)$, indicates that the contributions of the successive
eliminations of the degrees of freedom are piled up during the integration of
the differential equation and the solution of the renormalization group
equation resummes the perturbation series. The solution of the differential
equation interpolates between the bare and the effective potential
as $k$ is lowered from the original cut-off $\Lambda$ to zero.

\subsection{IR fixed point}
The infrared fixed point is always trivial for theories with mass gap.
In fact, as the block size extends beyond the correlation length
the evolution of the coupling constants slows down and we find a manifold
of stable fixed points. The only relevant coupling constants is the mass
since it is divergent in the units of the cut-off. But this controls a
quadratic operator so has trivial effects only. In order to find an
example where the IR fixed point generates non-renormalizable relevant
operators we look for models with massless excitations. The massless one
component $\phi^4$ model is not appropriate due to the Coleman-Weinberg
mechanism \cite{colwe}. We need a symmetry to keep the mass gap zero in the
presence of the interactions. The simplest example is the linear sigma model
in the symmetry broken phase. Let us start with the $O(N)$ invariant
lagrangian,
\be
L={1\over2}(\partial_\mu\phi^a)^2+U(|\phi^a|),
\ee
$a=1,\cdots,N$, in four dimensions.
The renormalization group equation for the potential is of the form \cite{sbn}
\be
k\partial_kU=-{k^4\over16\pi^2}\biggl(
\ln(k^2+\partial^2_\ell U)+(N-1)\ln(k^2+\partial^2_{tr}U)\biggr),
\ee
where $\partial_\ell$ and $\partial_{tr}$ denote the derivatives in
the along the vacuum expectation value and the transverse directions
of the internal space, respectively. The beta function of the n-order
massive mode vertex is defined by
\be
\beta_n=k\partial_k\partial^n_\ell U_k(\Phi).
\ee
Note that these functions depend explicitly on the scale $k$ and $\Phi$.
The strength of the effective interactions for the particlelike
excitations are obtained by setting $\Phi^a=<\phi^a(x)>$.

Let us identify the leading IR piece of the beta functions. Since
\be
\beta_1=-{k^4\over16\pi^2}\biggl(
{\partial^3_\ell U_k(\Phi)\over k^2+\partial^2_\ell U_k(\Phi)}+(N-1)
{\partial_\ell\partial^2_{tr}U_k(\Phi)\over k^2+\partial^2_{tr}U_k(\Phi)}
\biggr),
\ee
the most important IR contribution of the higher order beta functions comes
from the highest order power of the transverse denominator,
\be
\beta_n=(-1)^n(N-1){k^4\over16\pi^2}
\biggl({\partial_\ell\partial^2_{tr}U_k(\Phi)\over
k^2+\partial^2_{tr}U_k(\Phi)}\biggr)^n
(1+O(k^2/\partial_\ell\partial^2_{tr}U)).\label{irduvb}
\ee
The dimension of the corresponding coupling constant, $g_n$, is $4-n$ so the
beta function, $\tilde\beta_n$, for $\tilde g_n=g_nk^{n-4}$ in the IR regime is
\be
\tilde\beta_n=(-1)^n{N-1\over16\pi^2}
\biggl({\partial_\ell\partial^2_{tr}U_k(\Phi)\over
k+\partial^2_{tr}U_k(\Phi)/k}\biggr)^n
(1+O(k^2/\partial_\ell\partial^2_{tr}U))+(n-4)\tilde g_n.
\ee
According to the Goldstone theorem $\partial^2_{tr}U_{k=0}(<\phi>)=0$.
We shall assume that $\partial_\ell\partial^2_{tr}U_k(\Phi)>0$ and
$\partial^2_{tr}U_k(<\phi>)=o(k)$ which is supported by the simple
one-loop solution. The result is that the odd vertices whose beta function is
negative correspond to relevant coupling constants. The vertices
$\phi^n,\ n=6,8,\cdots$ which are irrelevant at the UV scaling regime
give rise relevant operator(s) at the IR fixed point of the symmetry broken
theory. The inclusion of the renormalization of $Z(\Phi)$ leaves our conclusion
unchanged \cite{sbn}.

\subsection{Couplings of the Goldstone modes}
The IR divergence in the beta functions poses an interesting problem
for the chiral perturbation expansion and the Standard Model.
It is well known that the effective theory of the Goldstone modes,
the nonlinear sigma model is IR finite. This comes about because the Goldstone
modes fluctuations have no restoring force by symmetry so they interact
with each other via gradient couplings which suppress the
IR divergences. This makes the chiral models which include only the
Goldstone modes IR finite.

It seems surprising at the first moment that the IR finiteness is lost
when the massive modes are added to the Goldstone particles. Though the
on-shell amplitudes remain IR finite off-shell the IR divergences appear
in the massive particle Green functions. These divergences can be found
in the one-loop effective potential of the linear sigma model where
the heavy particle legs are connected by a massless particle loop. This is
just the contribution singled out in (\ref{irduvb}). This divergence comes
from the coupling between the Goldstone and the heavy modes via the potential
$U(\phi^a)$. This coupling contains no derivatives and leads to the IR
divergence in the beta function. The absence of the derivatives in this
coupling can be understood by noting that the transverse modes of the
linear sigma model are not exactly the Goldstone modes but contain the
heavy mode at higher order in the fluctuations around the vacuum. Since
the massless propagator of Goldstone modes comes from the second order
contributions of the action and the mixing between the Goldstone and
the heavy modes is of higher order the massive mode interacts by itself
by the long range massless propagator.
This IR divergence of the effective potential
makes the dynamics of the condensate sensitive for the coupling constants
which would have been completely unimportant otherwise.
Similar strongly coupled long range interactions have been noticed
in the framework of our extended Nambu-Jona-Lasinio model, after
equation (\ref{enjle}).

In regard to the Standard Model this argument raises the possibility
that the higher order, non-renormalizable Higgs vertices which are
generated by the exchange of the superheavy particles of the GUT scale
may have an unusual strong influence on the low energy physics.
These vertices are certainly small as we follow the renormalized
trajectory around the energy scale of the Higgs mass, $M_H$, but
they may become large in the IR scaling regime. Since we have a differential
equation the initial conditions at the UV cut-off determine the effective
coupling
constants in the IR region. Both the suppression in the UV
and the amplification in the IR generate qualitatively similar power dependence
around the fixed points according to the linearized blocking equations.
So it seems plausible to expect seizable sensitivity of the IR physics on these
non-renromalizable bare parameters. If the detailed numerical solution
of the renormalization group equation reveals such a sensitivity
then new parameters of the Standard Model are found. These parameters
become important well below the mass gap. What degrees of freedom do they
influence ? The answer points again to the condensate since $<\phi(x)>$ is
determined by the long range interactions between the heavy and the Goldstone
modes. It is reasonable to assume that the number of new parameters is
the number of the dynamically generated condensate.
There is a difference with superconductivity in the manner the
non-renormalizable vertices influence the excitation spectrum. The four fermion
interaction leads to the Cooper pair formation whose condensate gives rise
a new low energy excitation spectrum. In the case of the Higgs particle
$<\phi(x)>$ generates the mass for the fermions and the gauge bosons.
Thus the modification of the strength of the condensate feeds back to the
whole mass spectrum of the theory.

\section{Conclusions}
The confinement of quarks has generated different models and led to
the development of a number of non-perturbative mechanisms. The present
lectures intend to introduce few new pieces to this collection.
These are the breakdown of the fundamental group symmetry,
the role of the nontrivial integration measure in the path integral,
the presence of singular configurations in renormalized theories
and the appearance of new parameters by the condensates or
the multiple fixed point structure. QCD has a lesson for us
to learn in each of these directions. Such a rich and many-sided
theory can only raise our determination to aim at synthesis
and arrive at a comprehensive understanding.

\section*{Acknowledgement}
I thank Laurent Lellouch and Vincenzo Branchina for the stimulating
discussions and the collaboration which gave a large part of the
material covered in these lectures.

\end{document}